# Degradation of alumina and zirconia toughened alumina (ZTA) hip prostheses tested under microseparation conditions in a shock device

*J. Uribe[1], J. Geringer[1,2,\*], L. Gremillard[3], B. Reynard[4]*

[1]Ecole Nationale Supérieure des Mines,
CIS-EMSE :UMR5146, LCG, F-42023 Saint-Etienne
[2]Penn State University, MSE-CEST, 206A Steidle Building,
University Park 16802 PA USA
\*corresponding author: geringer@emse.fr
[3]Université de Lyon, INSA-Lyon, UMR CNRS 5510 (MATEIS),
20 Avenue Albert Einstein, 69621 Villeurbanne Cedex, France
[4]Laboratoire de Sciences de la Terre UMR CNRS 5570
Ecole Normale Supérieure de Lyon
15 parvis René Descartes - BP 7000
69342 Lyon cedex 07, France

**Abstract**: This paper considers the degradation of alumina and zirconia toughened alumina vs. alumina for hip implants. The materials are as assumed to be load bearing surfaces subjected to shocks in wet conditions. The load is a peak of force; 9 kN was applied over 15 ms at 2 Hz for 800,000 cycles. The volumetric wear and roughness are lower for ZTA than for alumina. The long ZTA ageing did not seem to have a direct influence on the roughness. The ageing increased the wear volumes of ZTA and it was found to have a higher wear resistance compared to alumina.

## I.     Introduction

Total hip arthroplasty is continuously improving. New materials for hip joints must be submitted to gait biomechanics. When the hip joint is replaced, the ligament of the femoral head (round ligament) and the extracapsular ligaments are removed. Studies by video fluoroscopy have shown a separation of several millimeters between the cup and head during the swing phase [1,2]. Thus, when the heel touches the ground, a shock of up to 6 times the body weight occurs [3,4]. The microseparation during the gait cycle is shown in Figure 1 [5,6]. Materials should have good wear resistance and good shock resistance.

The bearing surfaces of artificial hip joints are classified into two types: Hard/Soft and Hard/Hard combinations. Metal-on-polyethylene is the more common of hard and soft pairs. This pair is very resistant to shock but less wear resistant, and debris could lead to an inflammatory response and eventually to aseptic loosening. Hard bearing surfaces such as Metal-on-Metal (MoM) have a high wear resistance, but the main disadvantage is the ion release, which has effects that are not well known. Ceramics, used as an alternative for the hard pairs, were investigated for Total Hip Arthroplasty, THA, in the beginning of 1970's. Boutin was the first orthopedic surgeon who implanted this kind of material in artificial hip joints [7-9].

Hip simulator studies, ISO 14242-1, demonstrate that combinations of ceramic on ceramic have a low wear rate compared to other pairs [10,11], which is an advantage. Nevertheless, ceramics are brittle, and some in vivo fractures have been reported. Specific wear zones were highlighted on the head of hip explants [12]. The actual process of degradation is related to the wear of alumina. The Low Temperature Degradation (LTD) of zirconia has been related to an increase of the roughness and the production of debris [13]. The Desmarquet's problem, which occurs with Prozyr® products, has led to huge problems concerning the implantation of zirconia in the 2000's [14]. Afterwards, alumina was the referenced ceramic material that was implanted as hip implants. Thus, much effort has been dedicated to developing new ceramic materials.

Thus, important improvements have been achieved in the biomaterials field, i.e., new ceramics with alumina matrix composites. Zirconia toughened alumina (ZTA) with small grains of zirconia has been developed with a $K_{IC}$ factor higher than those of common ceramics composed of alumina and zirconia ceramics [15,16]. Nevertheless, the phase transformation of zirconia grains remains a concern [17,18], and studies on avoiding low temperature degradation and ageing of zirconia have been performed [19].

In this paper, we study the resistance of alumina and ZTA ceramics subjected to shocks, i.e., impact loading of a few milliseconds, with a specific device. The influence of ZTA ageing on the wear resistance is also investigated.

Figure 1. Microseparation; distance (d) between the head and cup during the swing phase leading to a shock

**II. Materials and methods**

   *1. Device*

A special device was used (Quiri Hydromecanique). The movement of the head is in the vertical direction, the y-axis, and the cup is constrained to move only in the horizontal direction, the x-axis. The applied load was controlled by displacement, which gave more accurate forces than control by force [20]. The head is positioned on a cone composed of Ti-6Al-4V, and the cup is in a shell cemented to the metallic support. The head and cup were positioned with an angle of 45° to roughly reproduce the anatomical position (Figure 2).

Figure 2. Shock machine

*2. Materials and conditions of tests*

The materials tested were 12 ceramic pairs with a diameter 28 mm: 3 alumina balls against alumina cups and 9 zirconia toughened alumina (ZTA, 10 vol. % of zirconia in alumina matrix [19]) against alumina cups. The composition of alumina is in accordance with ISO 6474-1. The pristine microstructures of ZTA could be observed in [19]. It is significant that the zirconia content is under the percolation rate; zirconia grains are dispersed homogeneously in the alumina matrix. The mechanical properties of pure alumina, pure zirconia and ZTA are presented in Table 1.

| Material | $K_{IC}$ (MPa.$\sqrt{m}$) | $K_{I0}$ (MPa.$\sqrt{m}$) | Ultimate strength (MPa) | Hardness Vickers (MPa) |
|---|---|---|---|---|
| Alumina | 4.2 | 2.4 | 400-600 | 1800-2000 |
| Zirconia | 5.4 | 3.5 | 1000 | 1200-1300 |
| ZTA | 6.0 | 5.0 | 600 | 1800 |
| 12Ce-TZP | 7.8 | 5.1 | 700 | 1000-1100 |

Table 1. Mechanical properties of alumina, zirconia, zirconia toughened alumina, ZTA, and 12 mol. % $CeO_2$-doped Tetragonal Zirconia Polycrystals, 12Ce-TZP from [21].

For ZTA, the effect of ageing of the femoral heads was studied. According to [18,22] and ISO 6474-2, 10 hours of autoclaving at a temperature of 134°C and a pressure of 2 bars correspond to 40 years of in vivo ageing for ZTA of 17 vol. %. In this study, first, three femoral heads did not undergo ageing. Second, six heads composed of ZTA were artificially aged in an autoclave: 3 heads for 3 hours, i.e., approximately 12 years of actual aging. Finally, 3 heads were aged for 20 hours of actual ageing, i.e., approximately 80 years of ultimate ageing.
Tests related to shocks were carried out in wet conditions using newborn calf serum as a lubricant; the composition (calf serum heat inactivated, PAA, triple filtering steps at 0.1 µm, protein concentration of 17.5 g.L-1) is the same than the one suggested in ISO 14242-1 [23].
The test conditions are summarized in Table 2.
The 3D roughness, i.e., Sa parameter, measured by an optical profilometer (Bruker nanoscope, ex. Veeco, Wyko NT 9100) before the tests ranged between 28 and 34 nm for balls of alumina or ZTA regardless of the ageing time.
To test the ceramic materials in severe conditions, the load was a peak of force of 9 kN, which is equivalent to 6 times the body weight for a patient of 150 kg. The frequency was of 2 Hz, which is the double of the gait (Figure 3). As shown in Figure 3, the force is increased from 0 to 9 kN over roughly 15 ms.

| Materials (head/cup) | 3 alumina/alumina |
| --- | --- |
| | 3 ZTA/alumina |
| | 3 ZTA, aged 3 hours/alumina |
| | 3 ZTA, aged 20 hours/alumina |
| Test conditions | Wet (calf serum) |
| Load | Peak of 9 kN at 2 Hz |
| Microseparation | 1.3 mm |
| Test duration | 800,000 cycles |

Table 2. Experimental conditions

Figure 3. Scheme of the load applied during the gait and the load applied during the shock experiment; BW: Body Weight.

*3. Protocol for measuring wear volume and weight loss*

The wear of the heads was evaluated using two different techniques.

  a) Weight loss

Weight loss is a well known and simple method. The balls were weighed before and after the tests in a balance with a resolution of 0.1 mg after carefully drying and removing any debris. Gravimetric measures were taken as a reference for validating the developed profilometer protocol.

  b) Optical profilometry

The second technique used optical profilometry. The scheme of the protocol is shown in Fig 4. This method allows non-destructive analysis and is based on scanning white-light interferometry in which a spherical pattern of fringes is formed; finally, the height according to the peaks is measured.

To measure the depth of the wear zone, images over the whole width of the wear stripe are taken. The depth of worn area is calculated by comparing it to the unworn zone. Six images are taken for each wear stripe. The images are processed using the profilometer software (Vision 4.20 2002-2008 Veeco Instruments, Inc). A filter is then applied for the shape and to practice tilt removal. This filter depends on the topography of the worn surface profile. Each 2D profile is exported to be processed in Matlab®. The unworn profile, i.e., the original profile, of the worn area is reconstructed from the worn profile by fitting a quadratic function. It is possible to calculate the worn area by taking the arithmetic difference of both areas. For each image, the wear volume is calculated as the worn area multiplied by the width of the image, L.

Figure 4. Wear volume measured by optical profilometry.

*4. Phase transformation*

Phase transformation has been reported for retrieved femoral heads, and it has been related to the time of implantation. This phase transformation, from tetragonal to monoclinic, is studied by Raman spectroscopy (Horiba Jobin Yvon HR800, laser λ=514.5 nm) at latitudes of 90°, 45°, 0°, -45°. The method consists of measuring the intensity of the peaks: tetragonal ($I_{t1}$ and $I_{t2}$) and monoclinic ($I_{m1}$ and $I_{m2}$). The content of monoclinic phase is then calculated using the Clarke and Adar formula [24]:

$$\%monoclinic\_phase = \frac{I_m^{178} + I_m^{189}}{0.97(I_t^{145} + I_t^{256}) + I_m^{178} + I_m^{189}} \times 100 \qquad (1)$$

An example of a Raman spectrum for a ZTA ball is shown in Figure 5, and the content of the monoclinic phase is calculated as:

$$\% monoclinic = \frac{I_{m1} + I_{m2}}{0.97(I_{t1} + I_{t2}) + I_{m1} + I_{m2}} x100 = \frac{327 + 238}{0.97(243 + 451) + 327 + 238} x100 = 45.7\% \qquad (2)$$

Figure 5. Raman spectrum of a femoral head aged for 20 hours.

Moreover, scanning electron microscopy, JEOL JSM 6400, was practiced on the worn material.

**III. Results and discussion**

*1. Wear stripe location, width and roughness*

At the end of the test, all femoral heads exhibit two wear stripes (Figure 6). For ZTA, the lower wear stripe is no wider than 1 mm. For the alumina femoral heads, the width of the worn zone reached 3 mm. All stripes appear in the contact zone of the head and the rim of the cup near a latitude of 20°. The ratio between the worn and unworn area roughness is approximately 20 times for alumina and 2-3 times for ZTA.

The average roughness measured by profilometry is shown in Fig. 7. The alumina showed a roughness significantly higher than that of ZTA. Ageing of ZTA seems not to have a significant effect on roughness.

Figure 6. Location of wear stripes

Figure 7. Roughness of alumina and ZTA femoral heads

*2. Degradation mechanism/ images of the wear stripe*

Using the results of optical profilometry, the 3D images of the worn surfaces are shown in Figure 8. The ZTA surface is highly homogeneous. The worn area of alumina is significantly damaged, and it seems to have pits (Figure 8a). The wear stripes of alumina seem to be rougher than that of ZTA.

The SEM observations of the femoral heads show worn surfaces with grains that are pulled out of the alumina (Figure 9). For ZTA with 20 hours of ageing, the grains seem to be 'piled up', and the worn area appears to be covered by compacted debris. At the micron scale, the boundary between the worn and unworn areas was clearly distinguishable in the alumina heads, whereas, for the ZTA heads, the boundary is less well defined.

Figure 8. 3D profiles of a worn area before and after filtering: a) alumina and b) ZTA

Figure 9. SEM images of a) the boundary between worn and unworn areas and b) within the worn surface for alumina and ZTA femoral heads

*3. Phase transformation*

The influence of artificial ageing on the wear resistance of femoral heads is investigated. Three couples were tested: not aged, aged for 3 hours and aged for 20 hours. The influence of shocks is described: it is related to the phase transformation. The Raman spectra are taken before and after the test to determine the percent of the monoclinic phase for each step of ageing. The results are shown in Figure 10.

One way or two-way ANOVA tests (p = 0.05) demonstrate that there was no significant difference in the monoclinic content before and after testing. Moreover, they do not show a significant difference according to the duration of ageing. However, Student's t-test (p = 0.05) yielded comparison results between each sample. Each condition of ageing (not aged, 3 hours and 20 hours) is compared with another one, for a total of 14 combinations. First of all, for the sample after 20 hours of ageing, the percentage of monoclinic phase is significantly different in comparison with the sample aged for 3 hours and not aged. Second, femoral heads that are subjected to 3 hours and 20 hours of ageing show different percentages of monoclinic phase after degradation by shocks than before shocks. The surprising point is that the sample subjected to 20 hours of ageing has a lower percentage of monoclinic phase after shocks than before shocks. The discrepancy is greatest for this sample after shocks and after 20 hours of ageing. For the two other samples, i.e., the sample that was not aged and the sample subjected to 3 hours of ageing, shocks increase the percentage of the monoclinic phase. Thus, one might suggest that the crystallographic transformation is promoted by shocks.

Figure 10. Monoclinic content of ZTA femoral heads

*4. Weight loss and wear rate*

The results of the wear volume using both the weight loss and profilometry methods are shown in Figure 11. For alumina, the weight loss measurement is higher than the one related to profilometry. Both methods give similar values of the wear volume for ZTA. For ZTA aged 20 hours, the wear volume calculated by profilometry is much higher than weight loss. The choice of the best filter was studied. For alumina balls, a spherical filter is applied within the worn area. For ZTA, the spherical filter showed increased roughness and did not fit the surface shape. A cylindrical filter was then chosen. The different filters should explain the different ranking between weight loss and profilometry for the alumina head. This point needs further investigation to elucidate the precise role of the filtering process.

The classic weight loss protocol is the simpler and more widely used method for estimating the wear rate of ceramics. However, this technique presents some disadvantages. When the head is removed from the cone, there could be some weight loss due to wear between the cone and the head. The opposite case is also possible: debris accumulation generated by fretting corrosion might increase the mass of the head. Using optical profilometry, it is possible to measure only the wear due to the friction of both surfaces.

Figure 11. Wear volume for alumina and ZTA femoral heads after 0.8 million cycles

The wear of alumina/alumina tested *in vitro* is reported from 0.02 to 1.84 $mm^3$/ million cycles.

| Reference | Wear rate |
|---|---|
| Stewart et al. 2003 [24] | 0.2-1.84 $mm^3$/million cycles |
| Essner et al. 2005 [10] | 0.02-0.08 $mm^3$/million cycles |
| Nevelos et al. 2001 [25] | 1 $mm^3$/year |
| Walter et al. [26] | 0.7 $mm^3$/year |

Table 3. Wear rates for alumina from the literature

*5. Comparison of alumina and ZTA*

The results for alumina and ZTA are summarized in Table 4. The duration of ageing does not increase the surface roughness according to the ANOVA test ($p < 0.05$) for the worn and unworn zones of the head. The roughness of the worn zone of the ZTA composite is approximately 5 times lower than that of alumina. As mentioned for the surface roughness, ageing ZTA for 3 hours has no influence on the wear volume. The wear volume of ZTA is 33% lower than that of alumina. ZTA samples, as they are manufactured, can decrease the wear volume under shock degradation according to the experimental conditions of this study. This composite should be relevant for artificial hip joints in the future. No failure of the head occurred during the typical shock tests. The wear volume of ZTA is drastically lower than that of Ultra High Molecular Weight PolyEthylene, UHMWPE, which is used in Metal on Polymer pairs [27].

These results depend on the manufacturing process (sintering, polishing, etc.), especially for the cone of the head, which is important for the lifetime of the ceramic head. This type of ceramic could have good performance if this step is considered in manufacturing. No cup composed of ZTA was investigated with shock tests or manufactured in this study. For a Ceramic on Ceramic hip joint, cups should be manufactured and tested with a ZTA head for shock degradations. It would be fruitful to compare typical ceramic-ceramic (alumina/alumina) structures with the composites (ZTA/ZTA) available in the orthopedic field.

However, a relevant stage is missing in this study. The interactions between the ceramic debris due to shock degradations and bone cells (osteoblasts and/or osteoclasts) should be investigated. The first key point will be the separation of debris, including small particles of ceramic (the order of magnitude should be a few dozens of nanometers). Other points could be related to behavior of cells with debris. Some questions are under study: the impact of these debris on the apoptosis of cells, the inflammation responses, etc. These points will be investigated in the next experimental tests.

|  | Alumina | ZTA |
|---|---|---|
| **Number of shocks** | 800,000 | |
| **Wear stripe location** | Upper and lower | Upper and small lower wear stripe |
| **Width of wear stripe** | > 4 mm | 1.5 mm |
| *Roughness worn zone, Sa (nm)* | 325 | $69 \pm 16^{1, 2, 3}$ |
| *Roughness no worn zone, Sa (nm)* | 28 | $34 \pm 6^{1, 2, 3}$ |
| *Wear volume, profilometry $mm^3$/million cycles* | 0.18 | $0.12 \pm 0.01^{1, 2}$ |

Table 4. Results for alumina and ZTA; 1: not aged ZTA; 2: aged 3 hours; 3: aged 20 hours.

Finally, simulations by Finite Element Analysis should be performed to give more accurate information about the kinematics and maximum stresses on new design of heads and cups, for example. Moreover, the use of the friction law in finite element analysis has to be developed. A multi-scale analysis is in progress to characterize degradations of the composite under stresses close to the ones applied during shocks between ceramics [29].

### IV. Conclusions

The shock resistance of alumina and ZTA heads against alumina cups was investigated. The following conclusions can be drawn:

- The shock device reproduces the *in vivo* degradation mechanism. The wear stripe location and shape are comparable to cases reported in the literature. For alumina, the wear rate was found to be similar to the values of retrieved femoral heads.
  Alumina balls have wider wear stripes and a greater increase in roughness within the worn zone (20 times the initial value, i.e., no-worn zone) than ZTA (2-3 times the initial value). The 3D profilometric technique is used for estimating the wear volumes, and it is an original method. The results are promising. This method should be tested with other bearing surfaces for validation, such as Metal on Metal artificial hip joints.
- A phase transformation occurred according to the ageing process (Student's t-test) of ZTA heads. The ageing process has an impact on the percentage of the monoclinic phase; it increases according to the duration of ageing, and shocks promote the transformation.
- For wear induced by shocks, ZTA is more resistant than alumina in terms of wear volume. ZTA suffers degradations under no ageing and 3 hours of artificial ageing. For 20 hours of artificial ageing, the ultimate duration, ZTA seems to show weaknesses. In further investigations, experimental results for composites should be compared with retrieved femoral heads.

Numerous topics can be developed and investigated with this shock device: heads and cups composed of composites, comparison of wear volume results with explants (according to the schemes of manufacturers), elucidation of the role of ceramic debris on the bone cells and testing new couples, such as Metal on Metal hip joints.

**Acknowledgments**

This study was part of the program OPT-HIP. The authors acknowledge the Agence Nationale pour la Recherche, ANR, for financial support and N. Curt for technical collaboration with the shock device. In addition, the authors would like to acknowledge the Region Rhône-Alpes for the grant 'Explora Pro' at Penn State University, PA USA.